\documentclass[prd,aps,letterpaper,twocolumn,superscriptaddress,preprintnumbers,nofootinbib,floatfix]{revtex4}
\pdfoutput=1

\usepackage[pdftex]{graphicx}
\usepackage[latin1]{inputenc}
\usepackage{amsmath}
\usepackage{amsfonts}
\usepackage{amssymb}
\usepackage{rotating}
\usepackage{subfigure}
\usepackage{paralist} 
\usepackage{verbatim}
\usepackage{color}
\usepackage{soul} 
\usepackage{appendix}

\usepackage{natbib}
\usepackage{epsfig,hyperref}
\usepackage{array}
\usepackage{threeparttable}

\newcommand{\lvec}{\mathbf{l}}
\newcommand{\dtwol}[1]{\frac{d^2 \lvec_{#1}}{(2\pi)^2}}
\newcommand{\Lvec}{\mathbf{L}}
\newcommand{\phihat}{\hat{\phi}}
\newcommand{\nhat}{\hat{\mathbf{n}}}
\newcommand{\dtwolprime}{\frac{d^2 \lvec'}{(2\pi)^2}}

\usepackage{color}

\newcommand{\mnras}{Monthly Notices of the Royal Astronomical Society}
\newcommand{\apjl}{Astrophysical Journal Letters}

\newcommand{\aap}{Astronomy \& Astrophysics}
\newcommand{\jcap}{J. Cosmol. Astropart. Phys.}

\newcommand{\planck}{{\it{Planck~}}}

\usepackage{color}
\definecolor{orange}{rgb}{1,0.3,0}

\begin{document}

\title{Internal Delensing of Cosmic Microwave Background Acoustic Peaks}
\date{\today}

\author{Neelima Sehgal}
\affiliation{Physics and Astronomy Department, Stony Brook University, 
Stony Brook, NY 11794}
\author{Mathew S. Madhavacheril}
\affiliation{Physics and Astronomy Department, Stony Brook University, 
Stony Brook, NY 11794}
\affiliation{Department of Astrophysical Sciences, Princeton University, 
Princeton, NJ 08544}
\author{Blake Sherwin}
\affiliation{Berkeley Center for Cosmological Physics, LBL and
Department of Physics, University of California, Berkeley, CA, 94720}
\author{Alexander van Engelen}
\affiliation{Canadian Institute for Theoretical Astrophysics, University of
Toronto, Toronto, ON, Canada M5S 3H8}


\begin{abstract}
\vspace{0.5cm}
We present a method to delens the acoustic peaks of the CMB temperature and polarization power spectra internally, using lensing maps reconstructed from the CMB itself. We find that when delensing CMB acoustic peaks with a lensing potential map derived from the same CMB sky, a large bias arises in the delensed power spectrum.  The cause of this bias is that the noise in the reconstructed potential map is derived from, and hence correlated with, the CMB map when delensing. This bias is more significant relative to the signal than an analogous bias found when delensing CMB $B$ modes.  We calculate the leading term of this bias, which is present even in the absence of lensing.  We also demonstrate one method to remove this bias, using reconstructions from CMB angular scales within given ranges to delens CMB scales outside of those ranges.  Some details relevant for a realistic analysis are also discussed, such as the importance of removing mask-induced effects for successful delensing, and a useful null test, obtained from randomizing the phases of the reconstructed potential. Our findings should help current and next-generation CMB experiments obtain tighter parameter constraints via the internal removal of lensing-induced smoothing from temperature and $E$-mode acoustic peaks.    
\vspace{1cm}
\end{abstract}
\keywords{cosmic microwave background -- cosmology: observations -- gravitational lensing -- de-lensing -- inflation}

\maketitle

\section{Introduction}
\label{sec:intro}
\setcounter{footnote}{0} 

Cosmic Microwave Background (CMB) experiments have progressed from the first detections of gravitational lensing of the CMB by intervening large-scale structure \cite{smithlensing, hiratalensing, actlensing, engelenlensing, polarbearlensingA, polarbearlensingB, hansonBmode, storylensing, biceplensing} to a $40\sigma$ detection of the lensing potential power spectrum, corresponding to a measurement of its amplitude with $2.5\%$ precision \cite{plancklensing2013,plancklensing2015}.  These measurements have decreased the uncertainties on a variety of cosmological parameters \cite{actparams,planckParams2013,planckParams2015} as well as constrained the relationship between luminous and dark matter \cite{holdercib, planckciblensing, engelencib, allison, madhavacherillensing}.  Future measurements of the lensing of the CMB can be expected to yield powerful constraints on parameters such as the sum of the neutrino masses \cite{snowmassNus, cmbs4SB}.  At the same time, CMB lensing is also becoming a limiting source of noise for probing other fundamental physics.  For example, lensing of CMB polarization $E$ modes converts them into $B$ modes, which can obscure a primordial gravitational wave signal generated during inflation \cite{snowmassInf,cmbs4SB}.  In addition, the smoothing of the acoustic peaks of the temperature and $E$-mode power spectra due to lensing degrades the parameter constraints that can be achieved, such as on the number of light relic particles in the Universe \cite{cmbs4SB,baumannNeff,greenDelens}.  

The procedure to remove the lensing signal from CMB maps is called `delensing' \cite{knoxsongdelens, kesdendelens, seljakhiratadelens} and methods to achieve this with a reconstructed lensing field from the CMB itself or from a tracer of lensing such as the Cosmic Infrared Background (CIB) have been discussed in \cite{smithdelens, simarddelens, sherwindelens}.  Recently, \cite{LarsenPlanckDelens} have demonstrated delensing of the Planck temperature data using the CIB as a tracer of the lensing field.  As the sensitivity of CMB instruments improves, such as with the planned CMB-S4 experiment \cite{cmbs4SB}, delensing using a lensing field derived internally from CMB data, as opposed to an external tracer, will likely prove to be more powerful, since external tracers are not perfectly correlated with the underlying lensing potential \cite{planckciblensing, sherwindelens}.  Also recently, \cite{greenDelens} considered filtering schemes for delensing the acoustic peaks, and calculated the expected delensed power spectra and associated parameter constraints.  They also found that delensing always increases cosmological information and should thus be incorporated as part of the standard analysis for upcoming surveys.  That work was performed in an idealized context which did not take into account effects of the  finite survey region as well as effects  that arise if the lensing reconstruction is obtained from the same CMB modes that one is trying to delens.  In this work, we take a more  data-oriented approach and demonstrate a method to delens the CMB acoustic peaks using an internally-derived lensing field. In the process, we uncover and address these additional complications as they are present in a realistic analysis. We focus on delensing temperature maps as an example, but note that similar considerations apply for delensing $E$-mode acoustic peaks. The findings presented here will be important for next generation CMB experiments that will have the sensitivity to internally delens the acoustic peaks.  

In the next section, we outline the delensing pipeline employed.  In the subsequent section, we discuss a bias that arises if the lensing field is reconstructed using the same CMB that is being delensed, and show a method to avoid this bias.  We note that this bias is analogous to a similar bias found when delensing $B$ modes \cite{BmodeBias}, however, when delensing acoustic peaks, this bias is even more significant relative to the signal.  We also discuss the sensitivity of this delensing procedure to mask-induced effects, and a null test that can be used to cross check that delensing was performed successfully.  In the last section, we summarize and conclude.

\section{Delensing Pipeline}
\label{sec:pipeline}

\subsection{Simulations}
\label{sec:sims}
To explore delensing the temperature acoustic peaks we use 2000 simulations of lensed temperature maps, each with independent CMB and lensing potential realizations.  The simulations are generated as described in \cite{actlensing} and \cite{engelencib}. Gaussian-distributed primordial temperature maps and lensing potential maps are generated, and each CMB map is lensed with a potential map following the algorithm described in \cite{louislensing}.  A field corresponding to about 600 square degrees is then cut out of a larger CMB map and convolved with a 1.4 arcminute beam.  This mirrors the `D56' ACTPol field described in \cite{twoSeasonAuto}, as these two works share the same simulation set and some pipeline components.  However, we model the noise as $1\mu K$-arcmin white noise to correspond to forecasted levels for a CMB-S4 type experiment \cite{cmbs4SB}, and do not include an unresolved foreground component or atmospheric noise.  We have run this delensing pipeline on simulations with these extra noise sources and found the same behavior as we describe in the Results section.

\subsection{Lensing Reconstruction}
\label{sec:recon}
We obtain a reconstructed lensing potential map from a simulated lensed CMB map by exploiting the mode coupling that lensing induces.  Gravitational lensing by large-scale structure deflects the path of a CMB photon by an angle equal to the gradient of the projected gravitational potential, ${\bf{d}} = \nabla{\phi}$, where ${\bf{d}}$ is the deflection field, and $\phi$ is the projected potential given by  
\begin{equation}
\phi({\bf{\hat{n}}}) = -2 \int d\chi \frac{D_A(\chi_s-\chi)}{D_A(\chi)D_A(\chi_s)}\Psi(\chi{\bf{\hat{n}}},\chi).
\label{eq:phitheory}
\end{equation}
Here $\Psi({\bf{x}},\chi)$ is the three-dimensional gravitational potential, $\chi$ is the comoving coordinate distance, $\chi_s$ is the comoving coordinate distance to the last-scattering surface, and $D_A$ is the comoving angular diameter distance \cite{huCMBlens}, with $D_A(\chi) = \chi$ in a spatially flat universe. 
This deflection re-maps the primordial CMB and creates correlations between previously independent Fourier modes.

We use the optimal quadratic estimator \cite{huCMBlens, huokamoto} to isolate the lensing-induced mode coupling. This yields, in the flat-sky limit, an estimate of the lensing potential given by
\begin{equation}
\phihat(\Lvec) =  {A_{XY}(\Lvec)} \int \dtwol{} F^{XY}(\lvec, \Lvec -  \lvec) X(\lvec) Y(\Lvec - \lvec), 
\label{eq:phi}
\end{equation}
where $F$ is the filter function that optimizes the estimator, which in the case of $XY=TT$ is 
\begin{equation}
F_{TT}(\lvec_1,\lvec_2) = \frac{({\lvec_1 + \lvec_2})\cdot(\lvec_1  C^{TT}_{l_1}  + \lvec_2 C^{TT}_{l_2} )} {2(C^{TT}_{l_1} + N^{TT}_{\lvec_1})(C^{TT}_{l_2} + N^{TT}_{\lvec_2})}
\label{eq:bigF}
\end{equation} 
Here, $A_{XY}$ is a normalization function and $X$ and $Y$ represent any of $T$, $E$, or $B$ maps \cite{huokamoto}.  As a result, a lensing potential  estimate can be constructed from any pair of $T$, $E$, and $B$ or any of these paired with itself. In Eq.~\ref{eq:bigF}, $C^{TT}_l$ is the power spectrum including the peak smearing from lensing and $N^{TT}_\lvec$ is the noise power spectral density. 
 
As an example, we create a reconstructed lensing potential map from the CMB temperature map alone, using the $TT$ estimator.  For experiments with noise levels above $5\mu $K-arcmin, the $TT$ estimator will dominate the signal-to-noise of the reconstructed potential, so the following results are directly applicable.  For lower noise levels, the $EB$ estimator will dominate, however, non-zero correlations between the $EB$-reconstructed lensing map and the temperature and $E$-mode maps to be delensed, make the issues addressed here relevant in that case as well.  
 
To make the reconstructed lensing potential, we start with a simulated lensed temperature map convolved with a $1.4$ arcmin beam and with $1\mu K$-arcmin white noise.  We also make two splits of the lensed CMB temperature map that have a common signal and independent instrument noise realizations, with a noise level of $\sqrt{2}\times 1\mu K$-arcmin.  All three lensed CMB maps, the original and two splits, are apodized with a mask that has a cosine-squared edge roll-off with a width of 1.7 degrees.  Each of the three maps is then beam-deconvolved, and the non-split map is used for the lensing reconstruction.  For the reconstruction, we use only CMB modes between $l_\mathrm{min}=500$ and $l_{\rm{max}}=3000$ as a standard analysis would minimize bias from Galactic dust ($l_{\rm{min}}$ cut) and extragalactic foregrounds ($l_{\rm{max}}$ cut) \cite{vanengelenForegrounds, osborneForegrounds} while still maximizing signal-to-noise.  We also remove vertical and horizontal strips in the two-dimensional CMB Fourier-space map of $|l_x| < 90$ and $|l_y| < 50$, as is done in many ACTPol analyses, e.g.~\cite{louisPS, naessPS}. While we only consider using temperature maps to make the reconstructions here, we note that using polarization maps in addition should reduce the correlations between the noise in the reconstructed map and the temperature map to be delensed.  Using this reconstructed potential map, we delens the two splits of the lensed CMB map. We then take the cross spectra of these delensed splits to reduce instrumental noise bias, as in, e.g.,~\cite{louisPS}.  

When making the reconstructed lensing potential map, we calculate the normalization function $A_{XY}({\bf{L}})$ analytically following \cite{huokamoto}. To speed up the calculation in Eq.~\ref{eq:phi}, our pipeline writes the kernel of each estimator in a separable form as a sum of convolutions of two maps.  In this way, the convolutions can be calculated using multiplications in real space.  Since window functions induce mode coupling that affects the lensing estimator, we isolate this anisotropy signal from non-lensing mode couplings.  We call this non-lensing mode coupling the mean field, and subtract it from our lensing potential estimate.  The mean-field map is calculated by averaging the reconstructed potential obtained from Eq.~\ref{eq:phi} from our 2000 lensed CMB simulations, each with independent primary CMB and input lensing potential realizations.  Only the non-lensing-induced mode-coupling remains in the average, and subtracting this mean-field map from each reconstructed potential map results in an unbiased estimate of the potential:
\begin{equation}
\hat{\phi}^{\rm{unbiased}}_{XY}({\bf{L}}) = \hat{\phi}_{XY}({\bf{L}}) - \langle \hat{\phi}_{XY}({\bf{L}})\rangle
\label{eq:mf}
\end{equation}
The power spectrum of the mean-field map is larger than the true lensing potential power spectrum on large scales where $L \leq 100$.  Therefore, after subtracting this mean-field map from each reconstructed potential map, we in addition remove all scales with $L \leq 100$ from our resulting potential map and do not use them to delens the CMB map.  In practice when dealing with real data, we would make this $L$ cut to reduce bias in the reconstructed map that can arise from non-lensing induced mode couplings in the data that are not perfectly captured by the simulated mean field.  

\subsection{Delensing Procedure}
\label{sec:delensing}
Gravitational lensing shifts the unlensed CMB at position $\bf{\hat{n}}$, to a new position $\bf{\hat{n}+d}$, where the deflection angle is given by the gradient of the projected gravitational potential, $\nabla{\phi}$, and $\phi$ is as given in Eq.~\ref{eq:phitheory}, i.e.
\begin{equation}
T^{\rm{lensed}}({\bf{\hat{n}}}) = T^{\rm{unlensed}}({\bf{\hat{n}}}+\nabla{\phi}). 
\end{equation}
To delens the CMB we want to shift the positions of the lensed CMB by an estimate of $-\nabla{\phi}$ so that
\begin{equation}
T^{\rm{delensed}}({\bf{\hat{n}}}) = T^{\rm{lensed}}({\bf{\hat{n}}}-\nabla{\hat \phi}) 
\label{eq:delens}
\end{equation}
Following \cite{inverselens, greenDelens, LarsenPlanckDelens}, we evaluate $\nabla{\phi}$ at the displaced, as opposed to the unlensed position.  This is expected to be a good approximation because, while the deflections are on arcminute scales, they are correlated on scales of order degrees.
We Wiener filter our estimated $\phi$ map in Fourier space by multiplying Eq.~\ref{eq:phi} by $C_L^{\phi\phi}/(C_L^{\phi\phi} + A_{TT}(\Lvec))$ derived from theory, in order to down-weight the noisy modes.  We neglect additional filtering of the CMB $T$ field to select the imaged temperature modes, as advocated by Ref.~\cite{greenDelens}, because at our noise level all temperature modes are imaged above the noise.  We inverse Fourier transform the filtered $\hat{\phi}({\bf{L}})$ to get a $\hat\phi({\bf{\hat{n}}})$ map.  We then ``lens'' the lensed CMB map by the negative of the lensing potentail map, $-\nabla{\hat\phi}$, following the same algorithm in \cite{louislensing} used to generate lensed CMB maps from input $\nabla{\phi}$ maps.  Each lensed CMB split map is delensed in this way, and we then take the cross spectrum between them. To calculate the cross spectrum, we first compute the mode-coupling matrix as in, e.g.,~\cite{louisPS} to take into account the effects of the mask.  Applying the inverse of the mode-coupling matrix when taking the cross spectrum, we obtain the delensed CMB power spectrum.

\begin{figure}[t]
\hspace{0mm}\includegraphics[width=1.0\columnwidth]{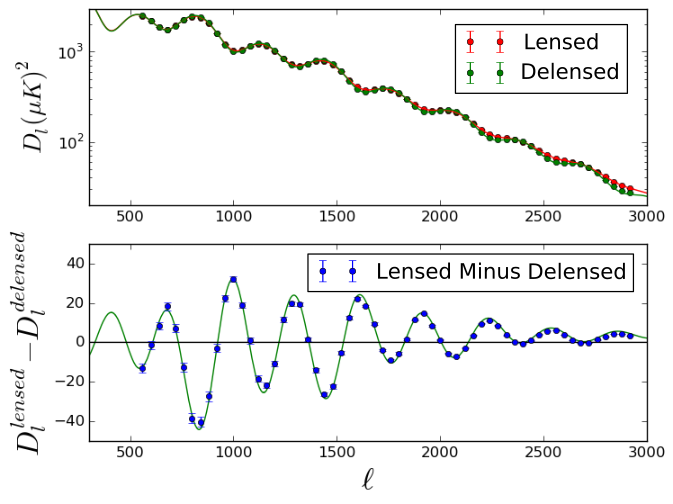}
\caption{{\it{Top panel:}} The mean temperature power spectra, $D_l=l(l+1)C_l/(2\pi)$, of lensed and delensed CMB maps from 2000 simulations are shown as red and green points, respectively. For this case, there is no noise added to simulations and the delensing is done with the input potential maps. {\it{Bottom panel:}} The mean lensed minus delensed power spectra from 2000 simulations are shown as blue points, and the errors represent the error on the mean.  The expected theory prediction of lensed minus unlensed power spectra is shown by the green curve. At high $\ell$, the blue points fall slightly below the green theory line due to the difference between delensing and inverse-lensing (see text for details).}
\vspace{0mm}
\label{fig:noNoise}
\end{figure}

\section{Results}
\label{sec:results}

In the top panel of Figure~\ref{fig:noNoise}, we show the power spectra of lensed and delensed CMB temperature maps in the case of no noise and perfect recovery of the input lensing potential.  We show the mean values from the 2000 lensed and delensed simulations (red and green points) as well as the lensed and unlensed input theory spectra (solid curves).  In the bottom panel of Figure~\ref{fig:noNoise}, we difference the lensed and delensed power from each simulation, and show the mean and errors on the mean.  We also show the expected theory prediction differencing the lensed and unlensed theory curves shown above.  In this no-noise case, where delensing is done with the input, as opposed to the reconstructed, potential map, one can near-perfectly remove the lensing-induced peak smearing from the power spectrum.  At high $\ell$, the delensed points fall slightly below the theory curve due to the difference between delensing and inverse-lensing.  Delensing (or anti-lensing) is what is done in this work: $T^{\rm{unlensed}}({\bf{\hat{n}}}) = T^{\rm{lensed}}({\bf{\hat{n}}}-\nabla{\phi})$, where $\nabla{\phi}$ is evaluated at the lensed, as opposed to the unlensed, position.  As can be seen from Figure~\ref{fig:noNoise}, this is a good approximation to inverse-lensing, which is the exact recovery of $T^{\rm{unlensed}}({\bf{\hat{n}}})$  using the unlensed position for evaluating $\nabla{\phi}$ \cite{inverselens, greenDelens,LarsenPlanckDelens}.

\subsection{Bias from Correlated Noise}

\begin{figure}[th!]
\hspace{0mm}\includegraphics[width=0.92\columnwidth]{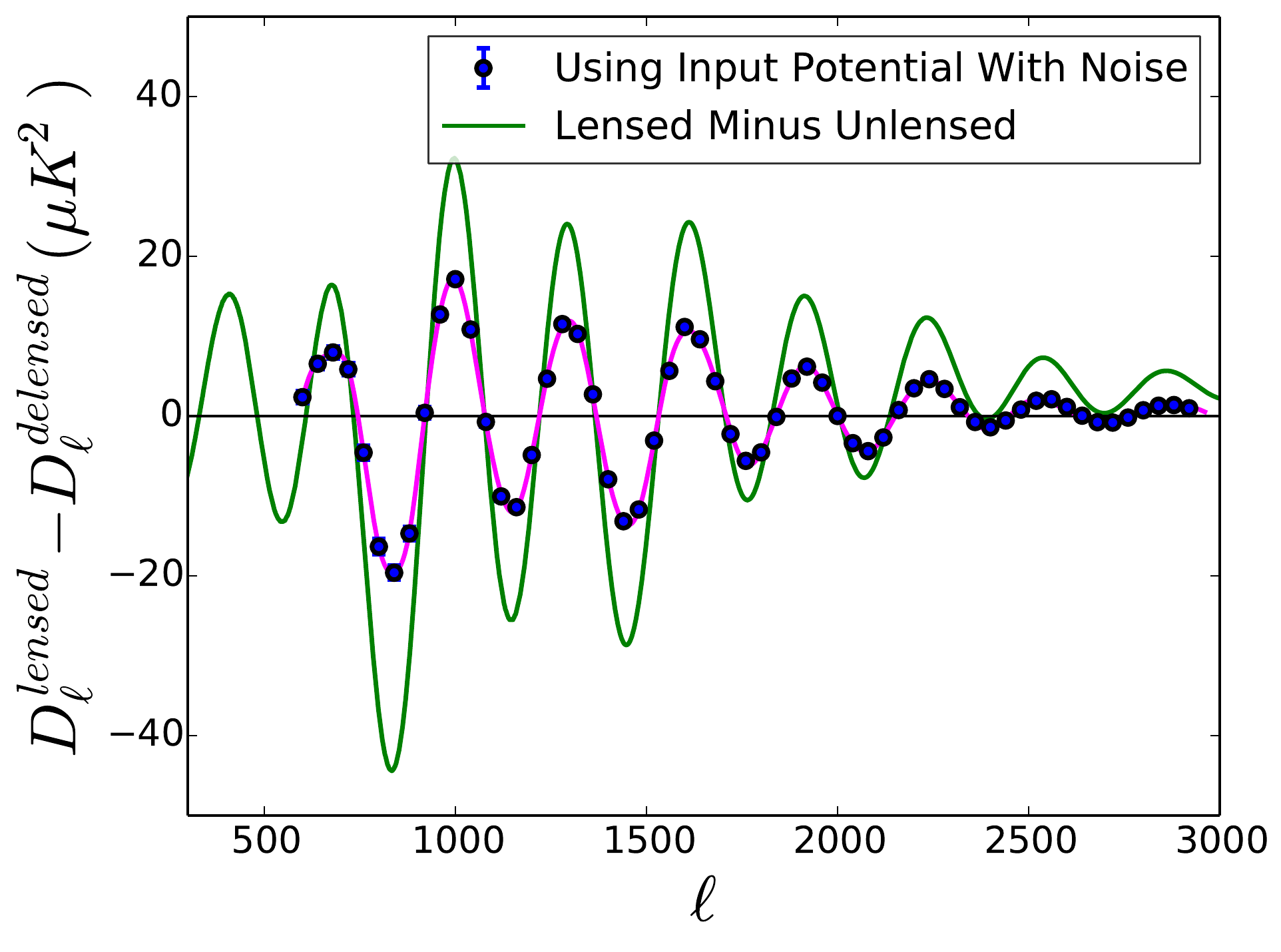}
\vskip -8.5mm
\hspace{0mm}\includegraphics[width=0.92\columnwidth]{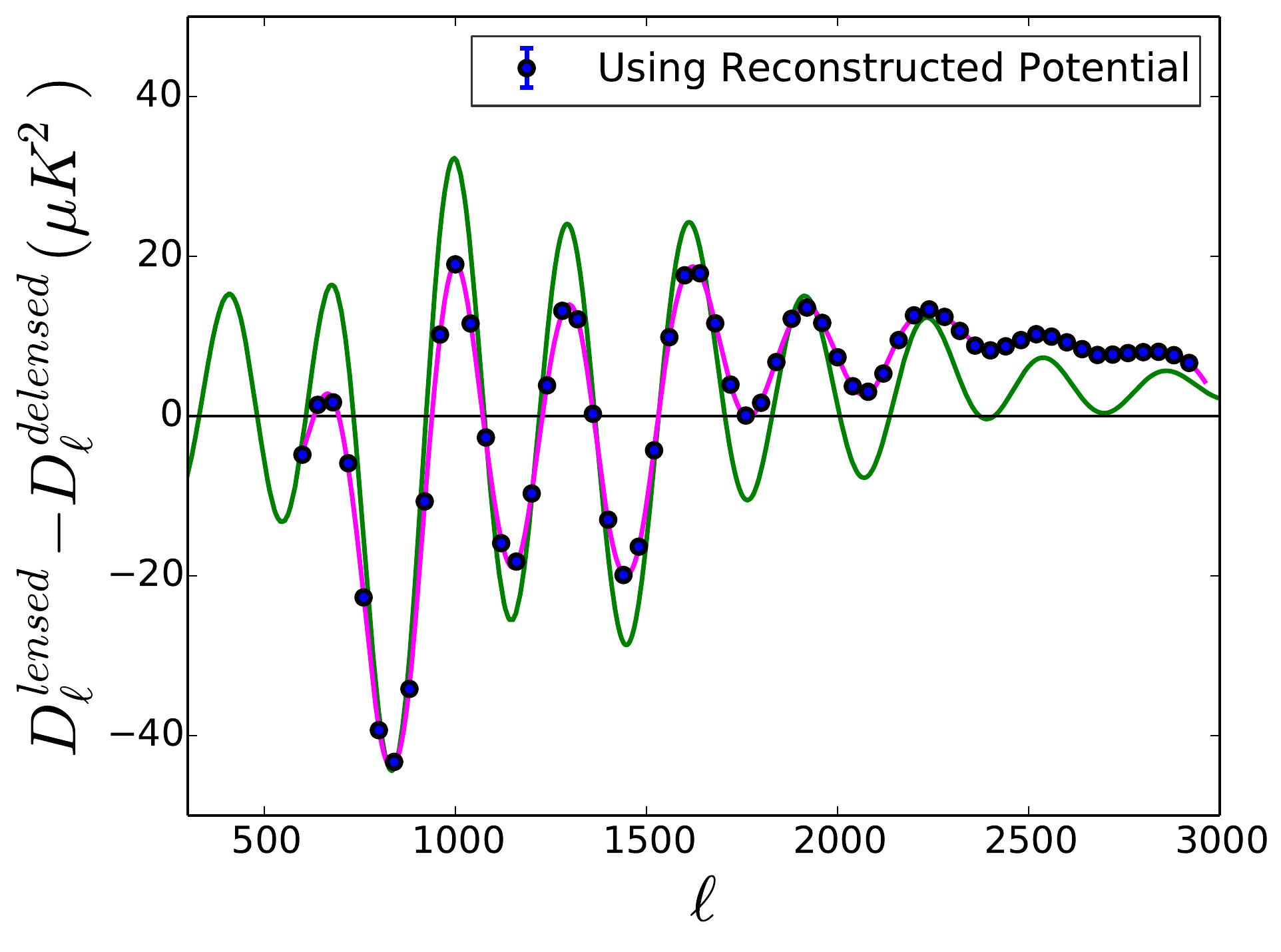}
\vskip -8.5mm
\hspace{0mm}\includegraphics[width=0.92\columnwidth]{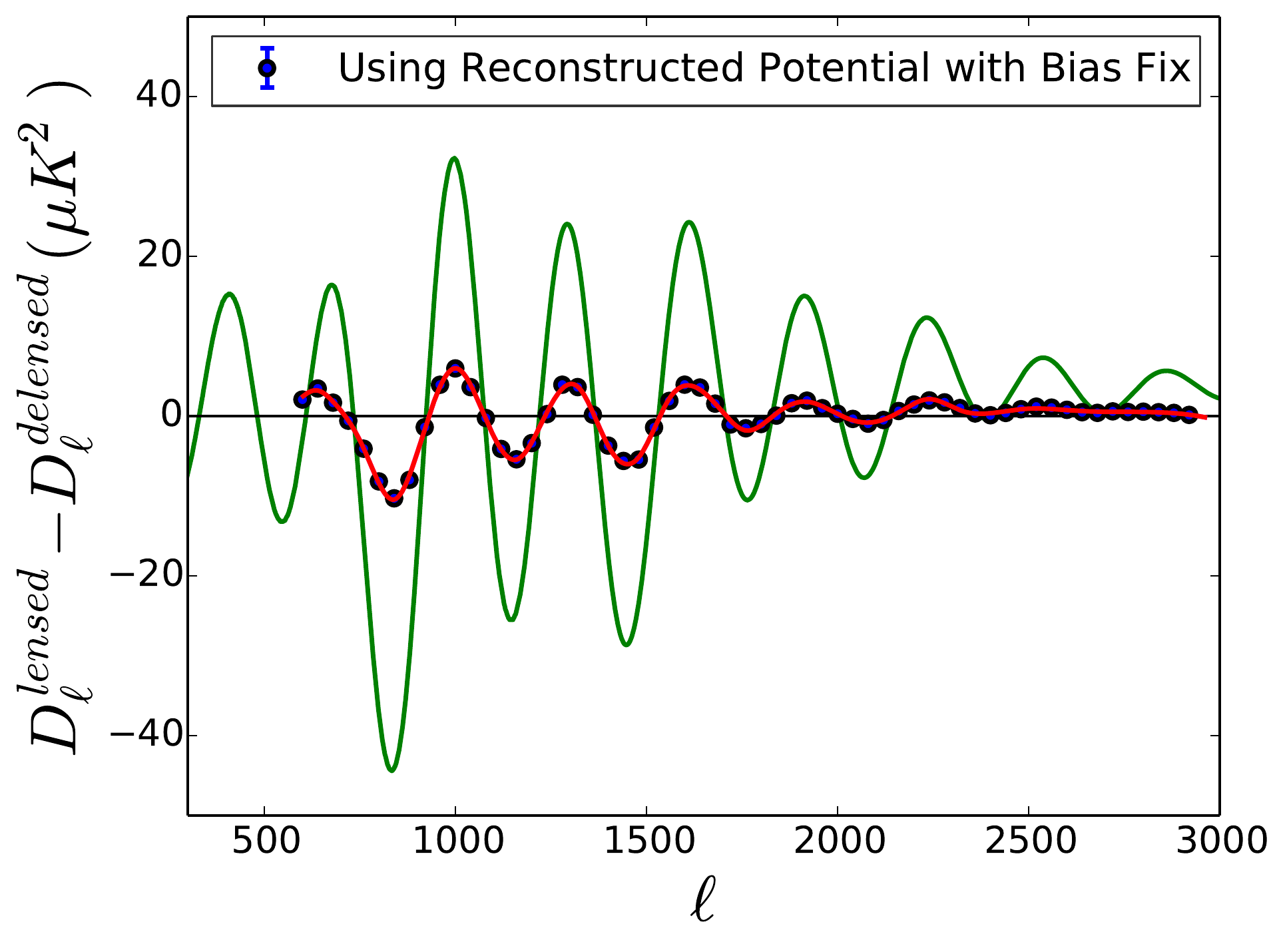}
\caption{{\it{Top panel:}} Blue points show lensed minus delensed power spectra when delensing with the input potential map with Gaussian reconstruction noise added corresponding to the expected noise when reconstructing from a CMB temperature map with $1\mu K$-arcmin white noise.  In this case, only some of the lensing signal is removed, as seen by the difference between the points and green curve, the latter of which shows the theory prediction of lensed minus unlensed power spectra. {\it{Middle panel:}} Same as top, but delensing with the reconstructed potential map.  The correlation between the reconstruction noise and the CMB map being delensed results in a large bias. {\it{Bottom panel:}}  Same as middle panel, except the bias has been removed by using the reconstruction from CMB $\ell$-modes in a given range to delens $\ell$-modes outside of that range. This procedure avoids correlated noise bias at the expense of some signal-to-noise, however more delensing can be recovered with modifications to this procedure (see text). }
\vspace{0mm}
\label{fig:bias}
\end{figure}

We now consider delensing in a more realistic experimental context.  In the top panel of Figure~\ref{fig:bias}, we again show the difference between the lensed and delensed power spectra when we delens with the input potential map, but we include lensing reconstruction noise corresponding to that expected with a CMB temperature map with $1\mu K$-arcmin white noise.  The reconstruction noise is from both the primary CMB and from the instrumental noise; we approximate it as Gaussian noise with the given power spectrum.  We see that even though we are using the input potential map to delens, the presence of reconstruction noise and our chosen cuts limit how much of the two-point lensing can be removed.  The error bars indicate the error on the mean from the 2000 simulations. We fit a smooth curve to the points shown with a cubic spline function.

Delensing the lensed CMB temperature map with $1\mu K$-arcmin noise using the reconstructed potential map derived from the same map yields the points shown in the middle panel of Figure~\ref{fig:bias}. 
Given that the noise power was the same as that in the upper panel, these points should have matched the points in the top panel of Figure~\ref{fig:bias}.  Instead they are significantly biased away from the top panel expectation.  This bias is due to the noise in the reconstructed potential map being correlated with the CMB map that is being delensed.  We note that any correlation between the CMB map being delensed and the CMB map used in the reconstruction of the potential map will result in some bias.  

Here we explain in more detail the origin of this bias.  Even if no lensing is present, if one performs the exercise of reconstructing the potential map with the same CMB map that is delensed, this bias will arise.  This can be verified by repeating the delensing procedure shown in the middle panel of Figure~\ref{fig:bias} with an unlensed CMB simulation, which yields a very similar result.  To see this mathematically, for the case of no lensing, we recall Eq.~\ref{eq:delens} and Taylor expand assuming small deflections, $\nabla{\phi}$, to get
\begin{eqnarray} \label{eq:lensmodecoupling}
T^\mathrm{delensed}(\nhat) & = & T(\nhat - \nabla \phihat(\nhat)) \nonumber \\
& = & T(\nhat) - \nabla T(\nhat) \cdot \nabla \phihat(\nhat) + \mathcal{O}(\phihat^2).
\label{eq:taylor}
\end{eqnarray}
At zeroth order in $\phi$, this process will lead to a bias term given by taking the Fourier transform of Eq.~\ref{eq:taylor}, replacing $\phi$ with an estimate of $\phi$ from a quadratic reconstruction, and then taking the power spectrum of that.  Thus Eq.~\ref{eq:taylor} becomes
\begin{equation}
T^\mathrm{delensed}(\lvec) = T(\lvec) + \int \dtwolprime \lvec^\prime \cdot (\lvec - \lvec^\prime) T(\lvec^\prime) \phi(\lvec - \lvec^\prime)
\label{eq:fourier}
\end{equation}
since the gradient becomes an $i{\bf{l}}$ and multiplication becomes convolution in Fourier space. 

Substituting the $\hat \phi$ of Eq.~\ref{eq:phi} into Eq.~\ref{eq:fourier}, and calculating the power spectrum yields
\begin{eqnarray} \label{eq:lensmodecoupling}
\langle T^\mathrm{delensed} (\lvec_a) T^\mathrm{delensed}(\lvec_b) \rangle \supset  \langle T(\lvec_a) T(\lvec_b) \rangle + \nonumber \\ 
2\int \dtwolprime \lvec^\prime \cdot (\lvec_b - \lvec^\prime) A_{TT}(\lvec^\prime)\int\dtwol{1}  F^{TT}(\lvec_1, \lvec^\prime -  \lvec_1) \nonumber \\ 
\times \langle T(\lvec_a) T(\lvec_b - \lvec^\prime) T(\lvec_1) T(\lvec^\prime - \lvec_1)   \rangle 
\end{eqnarray}
where $\supset$ indicates these are some of the components that make up the delensed power spectrum.  If we assume there is no lensing present in the temperature maps, and use Wick's theorem for a Gaussian random field\footnote{$\langle$ABCD$\rangle$= $\langle$AB$\rangle$$\langle$CD$\rangle$+ $\langle$AC$\rangle$$\langle$BD$\rangle$+ $\langle$AD$\rangle$$\langle$BC$\rangle$} together with the fact that $\langle T^{U}(\lvec_1)T^{U}(\lvec_2)\rangle = (2\pi)^2 \delta^2(\lvec_1+\lvec_2)C_{l_1}^U$, then we obtain
\begin{eqnarray} 
C_l^\mathrm{lensed} - C_l^\mathrm{delensed} \supset
4 
\int \dtwolprime \lvec^\prime \cdot (\lvec^\prime +  \lvec)
A_{TT}(\lvec^\prime)   \nonumber \\
\times  F^{TT}(-\lvec, \lvec^\prime +  \lvec) C_{l}C_{|\lvec + \lvec^\prime|}.
\label{eq:biasTerm}
\end{eqnarray}
The right side of Eq.~\ref{eq:biasTerm} is part of the bias shown in the middle panel of Figure~\ref{fig:bias}, and we note that higher order terms will give additional contributions.  

We remove this bias by splitting the lensed CMB temperature map into annuli in $\ell$-space, similar to what was done in \cite{BmodeBias} when delensing B-modes. We start at $\ell=500$ and end at $\ell=3000$, with each annulus having an $\ell$-width of 250.  For the first annulus, $\ell \in (500,750)$, we delens these $\ell$-modes using a reconstruction derived from  $\ell \in (800,3000)$. For the second annulus, $\ell \in (750,1000)$, we delens these modes using a reconstruction from $\ell$-modes in the range $\ell \in (500,700)$ and $\ell \in (1050,3000)$.  We repeat this for each annulus, delensing the $\ell$ modes in the annulus with a reconstruction from all $\ell$-modes excluding the ones in the annulus.  We also keep a buffer of $\ell$-width equal to 50 between the annulus $\ell$-range and the $\ell$-range used in the reconstruction to reduce bias from correlations between neighboring $\ell$-modes arising from the apodization window.  

\begin{figure}[t]
\hspace{0mm}\includegraphics[width=1.0\columnwidth]{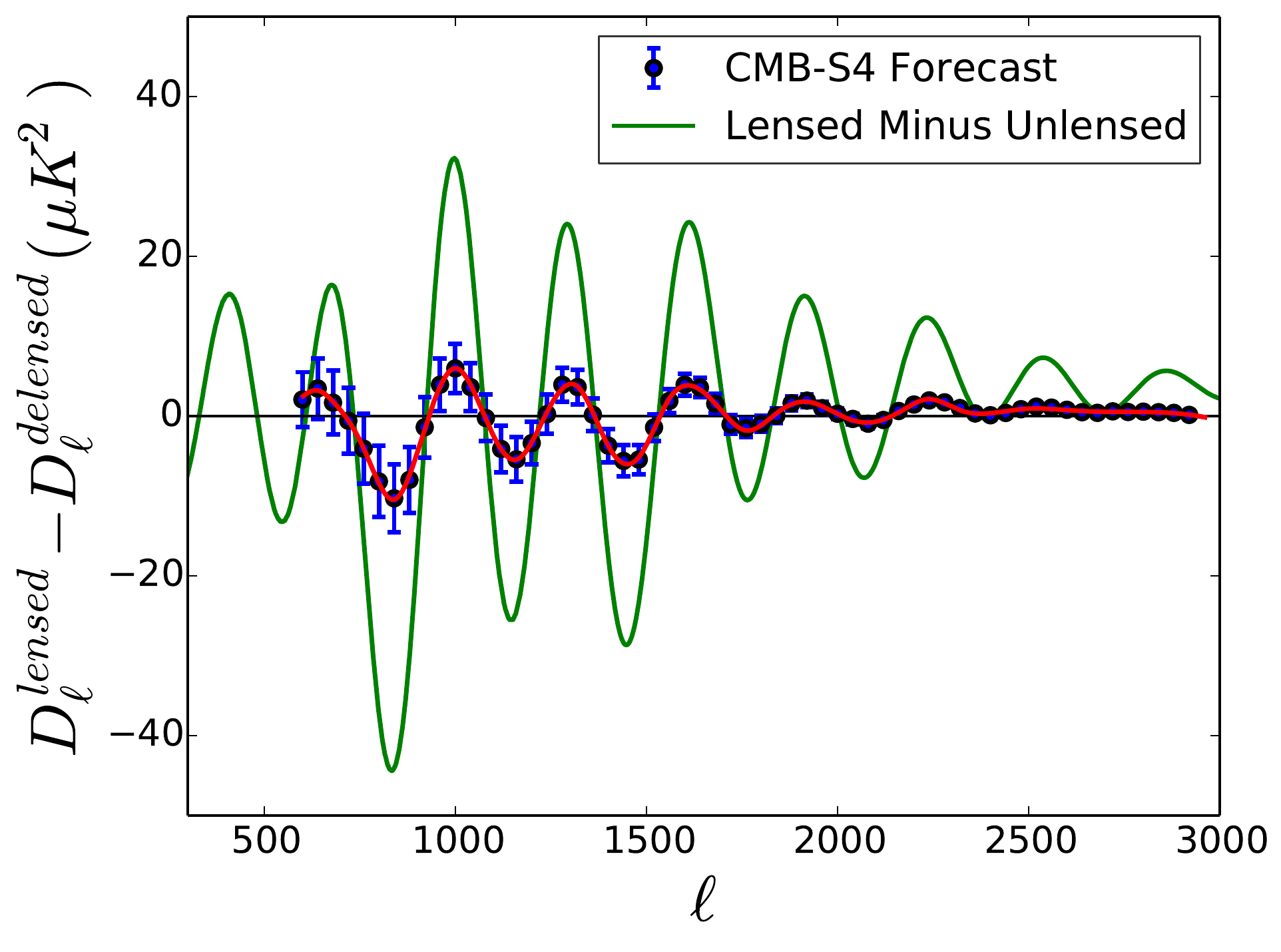}
\caption{Lensed minus delensed power spectra using a reconstructed potential plus bias fix as in the bottom panel of Figure~\ref{fig:bias}, now with error bars representative of a CMB-S4 type experiment with $f_{sky}=0.4$ and $1\mu K$-arcmin white noise.}
\vspace{0mm}
\label{fig:s4Forecast}
\end{figure}

The bottom panel of Figure~\ref{fig:bias} shows the result.  While this procedure eliminates the large bias shown in the panel above, less of the lensing-induced peak-smearing is removed as indicated by the lower amplitude of the points as compared to the curve in the top panel. In this case, the amplitude of the delensed curve is about a factor of 2 to 3 lower compared to the amplitude in the upper panel, although the relationship is not a direct scaling.  In terms of a detection of delensing, the bias-fix procedure results in a loss of signal-to-noise.  To quantify this loss for a CMB-S4 type experiment with $f_{sky}=0.4$ and $1\mu K$-arcmin white noise \cite{cmbs4SB}, we find that using an input potential plus noise would result in a $35\sigma$ detection, whereas using the reconstructed potential plus bias fix yields a $26\sigma$ detection, as shown in Figure~\ref{fig:s4Forecast}. 

In principle, one can split the CMB map into many more than 10 annuli, each with a smaller $\ell$-width, or one can also slice azimuthally \cite{sherwinBiasRemoval};  this would increase the signal-to-noise, as more modes would be used for each reconstruction.  However, keeping a non-zero $\ell$-width for the boundaries between each annulus and its inverse, as well as long computation times, will set practical limits for this procedure.  It should also be possible to estimate this bias term in a realization-dependent manner using the CMB map itself, for instance by evaluating Eq.~\ref{eq:biasTerm} with $C_l$ replaced by the power spectrum estimated directly from the data, using methods analogous to those introduced in \cite{dvorkinPatchytau, hanson2010, actlensing, namikawabias} for lensing power estimation. 

\subsection{Mean-field Subtraction}

\begin{figure}[t]
\hspace{0mm}\includegraphics[width=1.0\columnwidth]{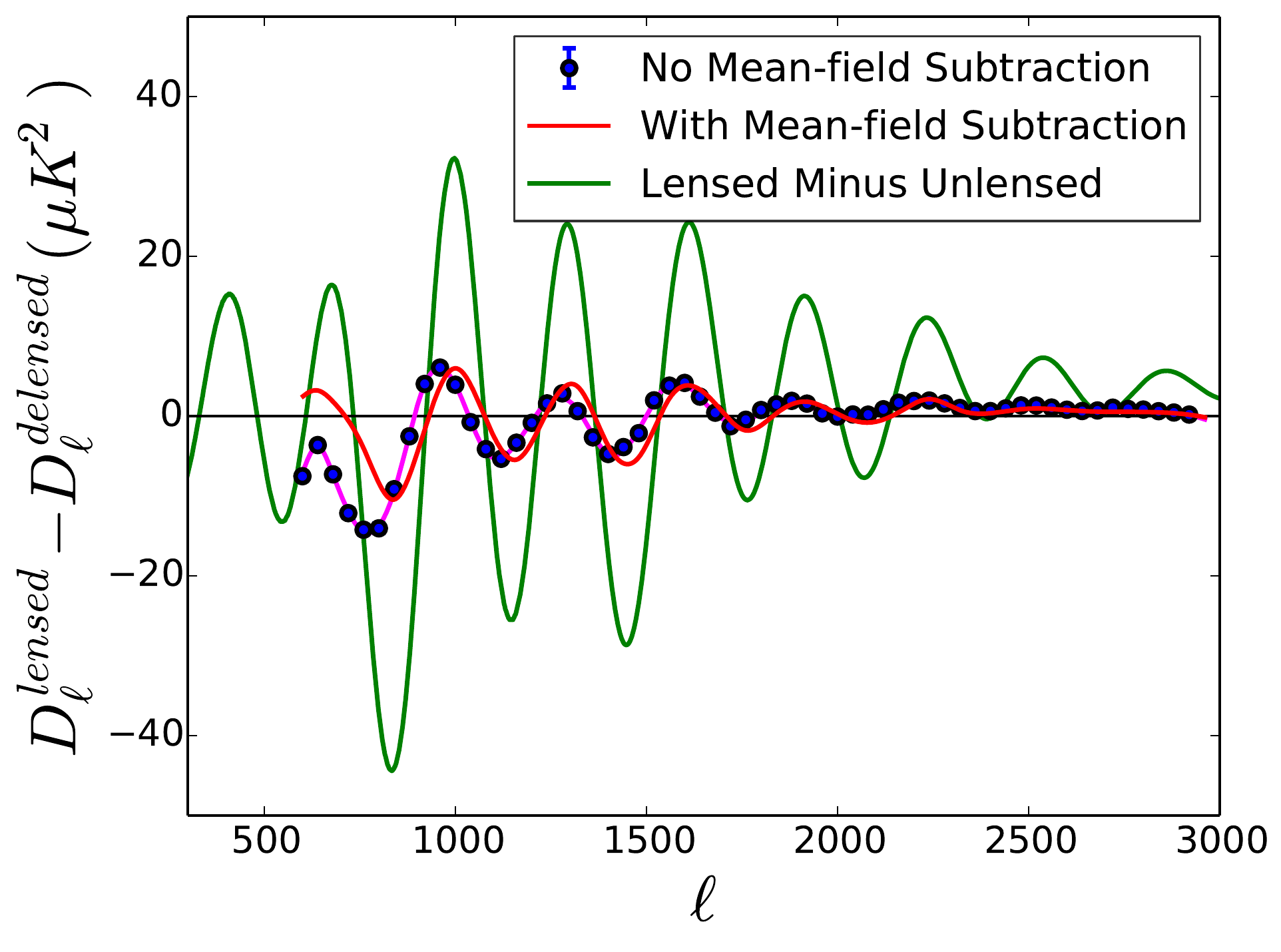}
\caption{The points show lensed minus delensed temperature maps when delensing with a reconstructed potential map that has not been mean-field subtracted.  The red curve shows the result from the bottom panel of Figure~\ref{fig:bias}, where the mean-field has been subtracted.  Even though modes with $L\leq 100$ have been removed in both cases, the significant difference between pink and red curves in this plot demonstrates the importance of mean-field subtraction for delensing acoustic peaks.}
\vspace{0mm}
\label{fig:meanfield}
\end{figure}

To delens the temperature and polarization acoustic peaks it is important to have simulations that represent the data so that the mean-field map can be properly modeled and subtracted from the reconstructed potential map.  Unlike when measuring the power spectrum of the lensing potential, where only the lowest $L$ bins are generally affected by inaccurate mean-field subtraction, a wide range of $\ell$-modes are improperly delensed as a result of inaccurate mean-field subtraction.  In Figure~\ref{fig:meanfield}, we show the result when the mean-field is not subtracted from the reconstructed potential map prior to using it to delens.  For comparison, we show the result when the mean-field has been subtracted, as reproduced from the bottom panel of Figure~\ref{fig:bias}.  Even though for both curves shown in this figure, the $L\leq 100$ modes have been removed prior to delensing, the significant difference between the curves shows how important accurately modeling and subtracting the mean field is for delensing the acoustic peaks. We also note that the offset in peak positions shown in Figure~\ref{fig:meanfield} between the delensed points with no mean-field subtraction and with proper mean-field subtraction suggests that this potentially can be used as a test to check for correct subtraction of the mean field.

\subsection{Null Test}

\begin{figure}[t]
\hspace{0mm}\includegraphics[width=1.0\columnwidth]{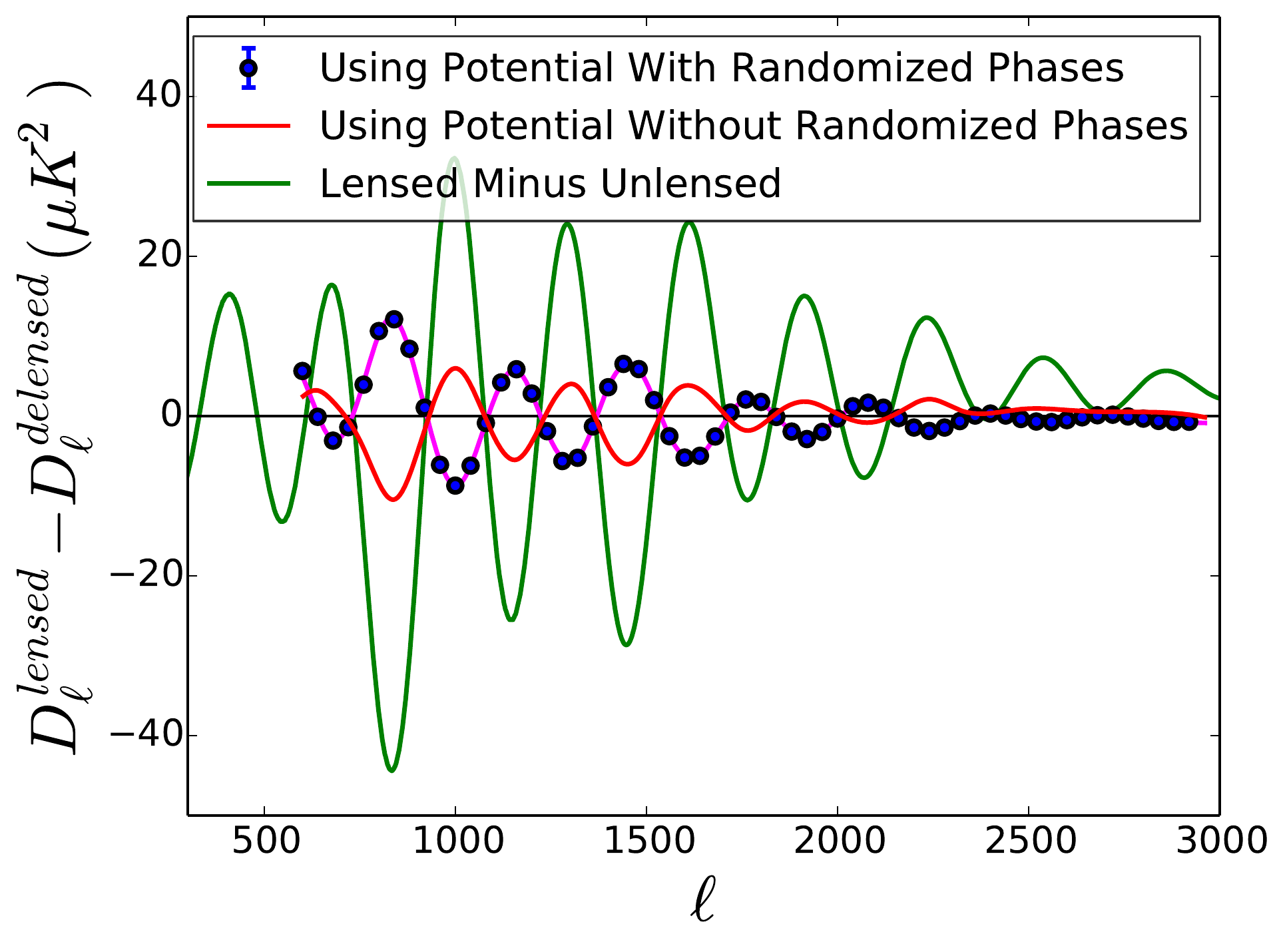}
\caption{The blue points show the power spectrum of lensed minus delensed temperature maps, when delensing with a reconstructed potential map whose phases have been randomized.  The peaks are now out of phase with the theory expectation for delensing with the correct potential map shown by the green curve in the case of no noise. Thus delensing with a reconstructed potential map where the phases have been randomized can serve as a null test to verify delensing is being done properly with the original reconstructed map.}
\vspace{0mm}
\label{fig:null}
\end{figure}

When CMB acoustic peaks are delensed with the correct lensing potential map, the difference between lensed and delensed spectra will have peaks and troughs aligned with the green theory curve in Figure~\ref{fig:null} that represents the difference between lensed and unlensed theory spectra.  However, when an incorrect potential map is used, instead of removing lensing from the lensed CMB map, one is effectively adding more lensing.  In this case, the difference between lensed and incorrectly delensed power spectra will be exactly out of phase with the theory expectation.  In Figure~\ref{fig:null}, we randomize the phases of the reconstructed potential map prior to delensing.  This effectively simulates delensing with an incorrect potential map, which has the same potential power as the correct map.  A similar test using an uncorrelated, simulated realization of the same lensing power was presented in \cite{LarsenPlanckDelens}.  Here we see the peaks and troughs of the delensed points are out of phase with the theory curve.  We consider this a null test in the sense that it demonstrates zero delensing.  Thus randomizing the phases of the reconstructed potential map prior to delensing, and checking that the peaks move out of phase with the expectation, can serve as a good null test to verify delensing is being done correctly. 

\section{Discussion}
\label{sec:discussion}

In this work, we presented a method to internally delens temperature acoustic peaks, which can easily be extended to delens $E$-mode acoustic peaks.  We identified a bias that results when the CMB map being delensed is also the same one used to reconstruct the potential.  The source of this bias is that the noise in the reconstructed map is correlated with the map being delensed.  We showed the source of the leading-order term of this bias mathematically, and demonstrated one method to remove it.  In addition, we addressed some practical considerations when doing a realistic analysis, such as the importance of carefully subtracting the mean-field, and presented a convenient null test.  The delensing pipeline presented here (potentially with an increase in the number of annuli) can be applied to current CMB datasets as well as near future CMB survey data from the Simons Observatory and CMB-S4.  This should allow improved constraints on interesting parameters such as the number of light relic species. 

\begin{acknowledgments}
\vspace{1cm}
NS would like to thank Thibaut Louis for useful discussions regarding the power spectrum analysis for masked, cut fields as well as for developing the code to lens flat-sky CMB simulations used in this work.  We thank Joel Meyers and Kendrick Smith for useful discussions, and the manuscript referee for useful suggestions.  NS acknowledges support from NSF grant number 1513618.   Computations were performed on the GPC supercomputer at the SciNet HPC Consortium. SciNet is funded by the CFI under the auspices of Compute Canada, the Government of Ontario, the Ontario Research Fund -- Research Excellence; and the University of Toronto.

\end{acknowledgments}


\begin{thebibliography}{3}

\bibitem[Smith et al.(2007)]{smithlensing} Smith, K.~M., Zahn, O., \& Dor{\'e}, O.\ 2007, \prd, 76, 043510 

\bibitem[Hirata et al.(2008)]{hiratalensing} Hirata, C.~M., Ho, S., Padmanabhan, N., Seljak, U., \& Bahcall, N.~A.\ 2008, \prd, 78, 043520 

\bibitem[Das et al.(2011)]{actlensing} Das, S., Sherwin, B.~D., Aguirre, P., et al.\ 2011, Physical Review Letters, 107, 021301 

\bibitem[van Engelen et al.(2012)]{engelenlensing} van Engelen, A., Keisler, R., Zahn, O., et al.\ 2012, \apj, 756, 142 

\bibitem[Ade et al.(2014)]{polarbearlensingA} Ade, P.~A.~R., Akiba, Y., Anthony, A.~E., et al.\ 2014, Physical Review Letters, 113, 021301 

\bibitem[Ade et al.(2014)]{polarbearlensingB} Ade, P.~A.~R., Akiba, Y., Anthony, A.~E., et al.\ 2014, Physical Review Letters, 112, 131302 

\bibitem[Hanson et al.(2013)]{hansonBmode} Hanson, D., Hoover, S., Crites, A., et al.\ 2013, Physical Review Letters, 111, 141301 

\bibitem[Story et al.(2015)]{storylensing} Story, K.~T., Hanson, D., Ade, P.~A.~R., et al.\ 2015, \apj, 810, 50 

\bibitem[Keck Array et al.(2016)]{biceplensing} Keck Array, T., BICEP2 Collaborations, :, et al.\ 2016, arXiv:1606.01968 

\bibitem[Planck Collaboration et al.(2014)]{plancklensing2013} Planck Collaboration, Ade, P.~A.~R., Aghanim, N., et al.\ 2014, \aap, 571, A17

\bibitem[Planck Collaboration et al.(2015)]{plancklensing2015} Planck Collaboration, Ade, P.~A.~R., Aghanim, N., et al.\ 2015, arXiv:1502.01591 

\bibitem[Sherwin et al.(2011)]{actparams} Sherwin, B.~D., Dunkley, J., Das, S., et al.\ 2011, Physical Review Letters, 107, 021302 

\bibitem[Planck Collaboration et al.(2014)]{planckParams2013} Planck Collaboration, Ade, P.~A.~R., Aghanim, N., et al.\ 2014, \aap, 571, A16 

\bibitem[Planck Collaboration et al.(2016)]{planckParams2015} Planck Collaboration, Ade, P.~A.~R., Aghanim, N., et al.\ 2016, \aap, 594, A13 

\bibitem[Holder et al.(2013)]{holdercib} Holder, G.~P., Viero, M.~P., Zahn, O., et al.\ 2013, \apjl, 771, LL16 

\bibitem[\planck Collaboration(2014b)]{planckciblensing} \planck Collaboration, Ade, P.~A.~R., Aghanim, N., et al.\ 2014a, \aap, 571, AA18 

\bibitem[van Engelen et al.(2014a)]{engelencib} van Engelen, A., Sherwin, B.~D., Sehgal, N., et al.\ 2014, arXiv:1412.0626 

\bibitem[Allison et al.(2015)]{allison} Allison, R., Lindsay, S.~N., Sherwin, B.~D., et al.\ 2015, \mnras, 451, 849 

\bibitem[Madhavacheril et al.(2015)]{madhavacherillensing} Madhavacheril, M., Sehgal, N., et al.\ 2015, Physical Review Letters, 114, 151302 

\bibitem[Abazajian et al.(2015)]{snowmassNus} Abazajian, K.~N., Arnold, K., Austermann, J., et al.\ 2015, Astroparticle Physics, 63, 66 

\bibitem[Abazajian et al.(2016)]{cmbs4SB} Abazajian, K.~N., Adshead, P., Ahmed, Z., et al.\ 2016, arXiv:1610.02743 

\bibitem[Abazajian et al.(2015)]{snowmassInf} Abazajian, K.~N., Arnold, K., Austermann, J., et al.\ 2015, Astroparticle Physics, 63, 55 

\bibitem[Green et al.(2016)]{greenDelens} Green, D., Meyers, J., \& van Engelen, A.\ 2016, arXiv:1609.08143 

\bibitem[Baumann et al.(2016)]{baumannNeff} Baumann, D., Green, D., Meyers, J., \& Wallisch, B.\ 2016, \jcap, 1, 007 

\bibitem[Knox \& Song(2002)]{knoxsongdelens} Knox, L., \& Song, Y.-S.\ 2002, Physical Review Letters, 89, 011303 

\bibitem[Kesden et al.(2002)]{kesdendelens} Kesden, M., Cooray, A., \& Kamionkowski, M.\ 2002, Physical Review Letters, 89, 011304 

\bibitem[Seljak \& Hirata(2004)]{seljakhiratadelens} Seljak, U., \& Hirata, C.~M.\ 2004, \prd, 69, 043005 

\bibitem[Smith et al.(2012)]{smithdelens} Smith, K.~M., Hanson, D., LoVerde, M., Hirata, C.~M., \& Zahn, O.\ 2012, \jcap, 6, 014 

\bibitem[Simard et al.(2015)]{simarddelens} Simard, G., Hanson, D., \& Holder, G.\ 2015, \apj, 807, 166 

\bibitem[Sherwin \& Schmittfull(2015)]{sherwindelens} Sherwin, B.~D., \& Schmittfull, M.\ 2015, \prd, 92, 043005 

\bibitem[Larsen et al.(2016)]{LarsenPlanckDelens} Larsen, P., Challinor, A., Sherwin, B.~D., \& Mak, D.\ 2016, Physical Review Letters, 117, 151102 

\bibitem[Teng et al.(2011)]{BmodeBias} Teng, W.-H., Kuo, C.-L., \& Proty Wu, J.-H.\ 2011, arXiv:1102.5729 

\bibitem[Louis et al.(2013)]{louislensing} Louis, T., N{\ae}ss, S., Das, S., Dunkley, J., \& Sherwin, B.\ 2013, \mnras, 435, 2040 

\bibitem[Sherwin et al.(2016)]{twoSeasonAuto} Sherwin, B.~D., van Engelen, A., Sehgal, N., et al.\ 2016, arXiv:1611.09753

\bibitem[Hu(2001)]{huCMBlens} Hu, W.\ 2001, \apjl, 557, L79 

\bibitem[Hu \& Okamoto(2002)]{huokamoto} Hu, W., \& Okamoto, T.\ 2002, \apj, 574, 566 

\bibitem[Naess et al.(2014)]{naessPS} Naess, S., Hasselfield, M., McMahon, J., et al.\ 2014, \jcap, 10, 007 

\bibitem[Anderes et al.(2015)]{inverselens} Anderes, E., Wandelt, B.~D., \& Lavaux, G.\ 2015, \apj, 808, 152 



\bibitem[van Engelen et al.(2014)]{vanengelenForegrounds} van Engelen, A., Bhattacharya, S., Sehgal, N., et al.\ 2014, \apj, 786, 13 

\bibitem[Osborne et al.(2014)]{osborneForegrounds} Osborne, S.~J., Hanson, D., \& Dor{\'e}, O.\ 2014, \jcap, 3, 024 





\bibitem[Louis et al.(2016)]{louisPS} Louis, T., Grace, E., Hasselfield, M., et al.\ 2016, arXiv:1610.02360 

\bibitem[Sherwin \& Das(2010)]{sherwinBiasRemoval} Sherwin, B.~D., \& Das, S.\ 2010, arXiv:1011.4510 

\bibitem[Dvorkin \& Smith(2009)]{dvorkinPatchytau} Dvorkin, C., \& Smith, K.~M.\ 2009, \prd, 79, 043003 

\bibitem[Hanson et al.(2011)]{hanson2010} Hanson, D., Challinor, A., Efstathiou, G., \& Bielewicz, P.\ 2011, \prd, 83, 043005 

\bibitem[Namikawa et al.(2013)]{namikawabias} Namikawa, T., Hanson, D., \& Takahashi, R.\ 2013, \mnras, 431, 609 

\end{thebibliography}
\end{document}